\documentclass[11pt]{article}
\usepackage{moriond,epsfig}

\usepackage{url}

\def\be{\begin{equation}}
\def\ee{\end{equation}}
\def\bea{\begin{eqnarray}}
\def\eea{\end{eqnarray}}

\begin{document}
\vspace*{4cm} \title{NA61-SHINE: HADRON PRODUCTION MEASUREMENTS FOR
  COSMIC RAY AND NEUTRINO EXPERIMENTS.}

\author{ N.~ABGRALL\\
on behalf of the NA61 and T2K collaborations. }

\address{D\'epartement de Physique Nucl\'eaire et Corpusculaire -
  DPNC,
  24 quai Ernest Ansermet,\\
  1205 Gen\`eve, Switzerland}

\maketitle\abstracts{ As neutrino long baseline experiments enter a
  new domain of precision, important systematic errors due to poor
  knowledge of production cross-sections for pions and kaons require
  more dedicated measurements for precise neutrino flux
  predictions. The cosmic ray experiments require dedicated hadron
  production measurements to tune simulation models used to describe
  air shower profiles. Among other goals, the NA61-SHINE~\cite{NA61}
  (SPS Heavy Ion and Neutrino Experiment) experiment at the CERN SPS
  aims at precision measurements (5\% and below) for both neutrino and
  cosmic ray experiments: those will improve the prediction of the
  neutrino flux for the T2K~\cite{T2K} experiment at J-PARC and the
  prediction of muon production in the propagation of air showers for
  the Auger~\cite{Auger} and KASCADE~\cite{KASCADE}
  experiments. Motivations for new hadron production measurements are
  briefly discussed. NA61-SHINE took data during a pilot run in 2007
  and in 2009 with different Carbon targets. The NA61-SHINE setup and
  preliminary spectra for positive and negative pions obtained with
  the 2007 thin (4\% interaction length) Carbon target data are
  presented. The use of the NA61 data for the T2K neutrino flux
  predictions is finally discussed in further details.}

\section{Needs for new hadron production measurements}
Many hadron production experiments have been conducted over a range of
incident proton momenta from 3 GeV/c to 450 GeV/c. However, most of
them cover only limited ranges in
momentum $p$ and production angle $\theta$
(or Feynman scaling variable $x_F$ and transverse momentum $p_T$).\\
Several models of secondary production have been derived by fitting
and interpolating experimental data on $p+A\rightarrow\pi^{\pm}X$ or
$p+A\rightarrow KX$. Shower cascade models (e.g. MARS~\cite{MARS},
FLUKA~\cite{FLUKA}) contain a number of physical assumptions and
cannot be modified by users. Parametrizations
(e.g. Sanford-Wang~\cite{S-W}, Malensek~\cite{Malensek}) account for
various aspects of production cross-sections such as $p_T$-scale
breaking but do depend
on the nuclear target properties, re-interactions, etc.\\
The lack of hadron production data requires reliance on such models to
extrapolate from existing data to the conditions of a given
experiment.
These extrapolations imply large and poorly known systematic
uncertainties.  Muon and neutrino flux predictions for current and
projected cosmic ray and neutrino experiments will require a
precision better than that obtained from those extrapolations. New
hadron production data at required projectile momentum and with
relevant targets are mandatory to reach the goals of those
experiments.

\section{The NA61-SHINE measurements}
\subsection{The NA61-SHINE experimental setup}
\begin{figure}[!h]
\begin{center}
\epsfig{figure=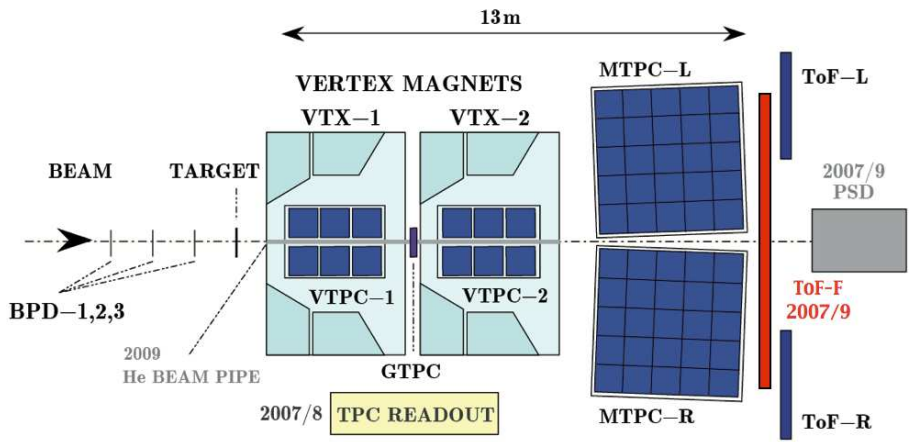, width=0.75\textwidth}
\end{center}
\caption{The NA61-SHINE apparatus. Dates refer to different
  installation periods and upgrades.}
\label{fig:apparatus}
\end{figure}

The NA61-SHINE apparatus~\cite{NA61-status} (see Fig.~\ref{fig:apparatus}) is a large
acceptance spectrometer which consists in a set of five time
projection chambers (TPCs): two TPCs, referred to as vertex TPCs, are
embedded in dipole magnets (1 Tm) and provide a high momentum
resolution, while two larger TPCs (main TPCs) are placed downstream
out of the magnetic field region. A smaller TPC, referred to as GAP
TPC, is placed in between the two vertex TPCs. This set of TPCs is complemented by
an upgraded time-of-flight ($ToF$) system with 120 and 70 ps resolution
for the forward and left/right walls respectively.

\begin{figure}[!h]
\begin{center}
\epsfig{figure=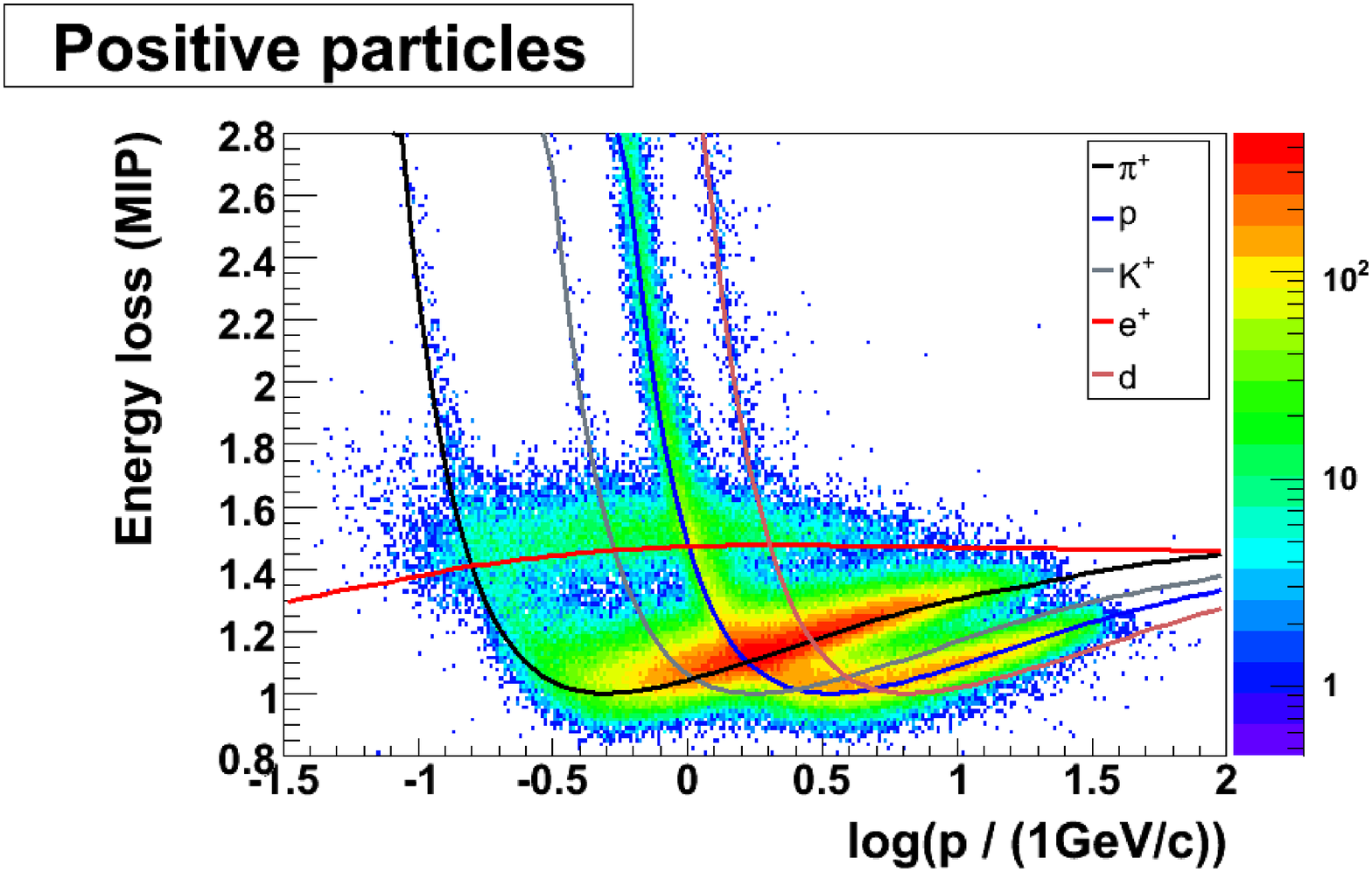, width=0.5\textwidth}
\epsfig{figure=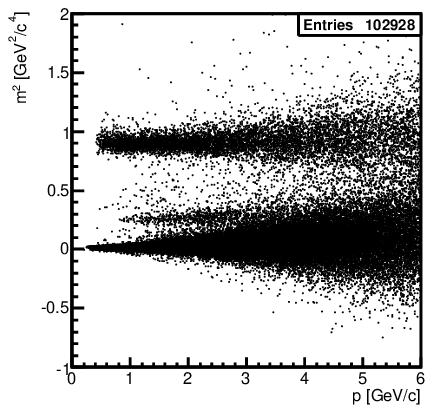, width=0.3\textwidth}
\epsfig{figure=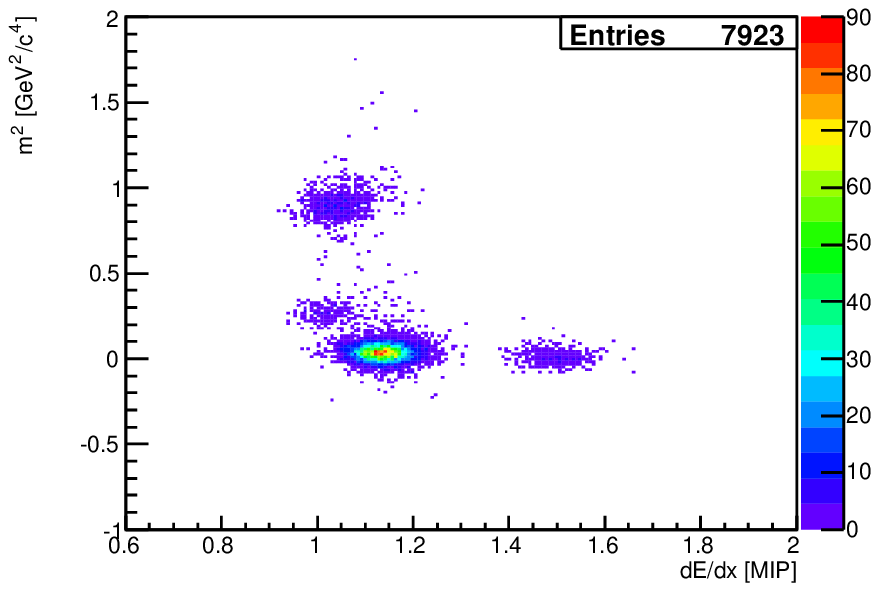, width=0.4\textwidth}
\end{center}
\caption{Energy loss versus momentum (left)
  and mass squared versus momentum spectra (right) for positive
  particles. Combined energy loss and time-of-flight measurements for
  all particles in the momentum range [2.,2.5] GeV/c (bottom).}
\label{fig:measurements}
\end{figure}

\begin{figure}[!h]
\begin{center}
\epsfig{figure=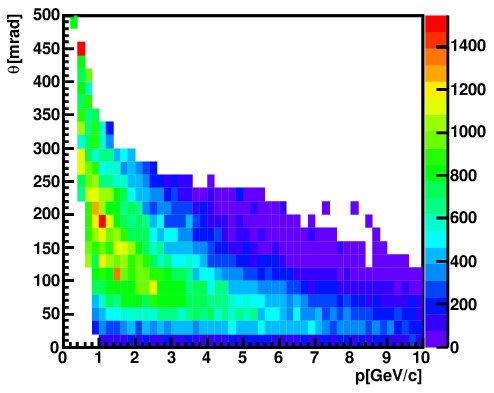, width=0.41\textwidth}
\epsfig{figure=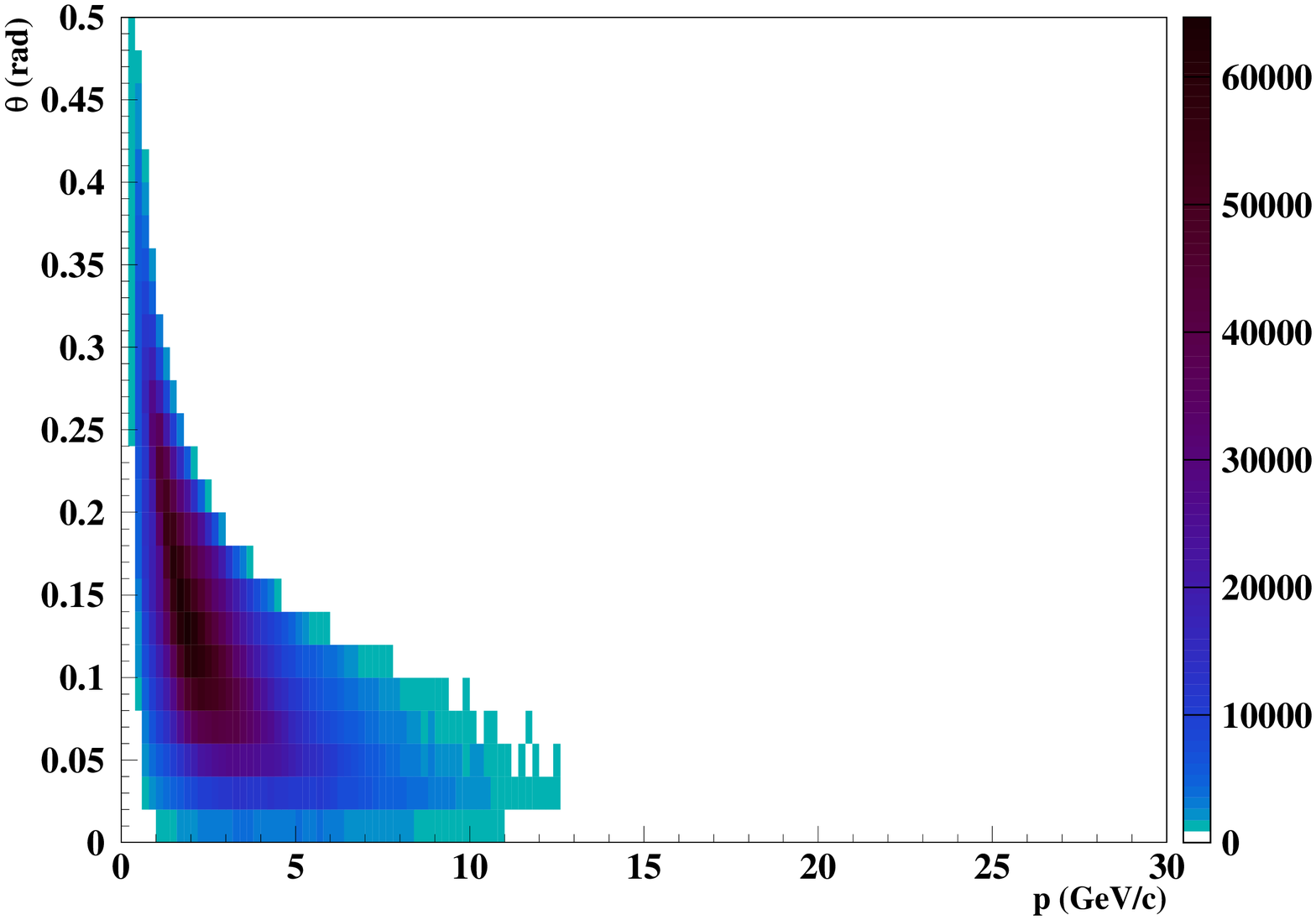, width=0.48\textwidth}
\end{center}
\caption{Absolute $\pi^+$ spectrum in the \{$p,\theta$\} phase space
measured in NA61 (left). Distribution of $\pi^+$'s producing neutrinos
in the far detector of the T2K experiment.}
\label{fig:acc-t2k}
\end{figure}

\noindent The NA61-SHINE apparatus provides high quality measurements
of both energy loss ($dE/dx$) and time of flight (see
Fig.~\ref{fig:measurements}). These measurements allow for particle
identification over a large range of momentum: $dE/dx$ alone is used
to identify particles below 1 GeV/c and in the relativistic rise
region of the Bethe-Bloch curves, while time-of-flight alone can be
used between 1 and 6 GeV/c. The combination of both measurements
provides a powerful separation of the
different particle species over a wider momentum range.\\
Data were taken in 2007~\cite{NA61-status} and again in 2009 after a
major readout upgrade for incoming protons of 31 GeV/c momentum
(corresponding to the T2K beam momentum), using both a thin Carbon
target (4\% of interaction length) and a full size replica of the T2K
target (1.9 interaction length). For the cosmic ray program, data were
taken for
incoming pions of 158 and 350 GeV/c momentum.\\
The large acceptance of the NA61-SHINE apparatus covers the relevant
phase space of both T2K and Auger experiments. As an example,
Fig.~\ref{fig:acc-t2k} compares the absolute (corrected) $\pi^+$
distribution in the \{$p,\theta$\} ($\theta$ is the production angle
with respect to the beam direction) phase space measured by NA61 with
the thin Carbon target and that of $\pi^+$'s from the primary
interaction producing neutrinos in the far detector of the T2K
experiment obtained from the T2K beam simulation.

\subsection{Preliminary spectra for positive and negative pions}
\begin{figure}[!h]
\begin{center}
\epsfig{figure=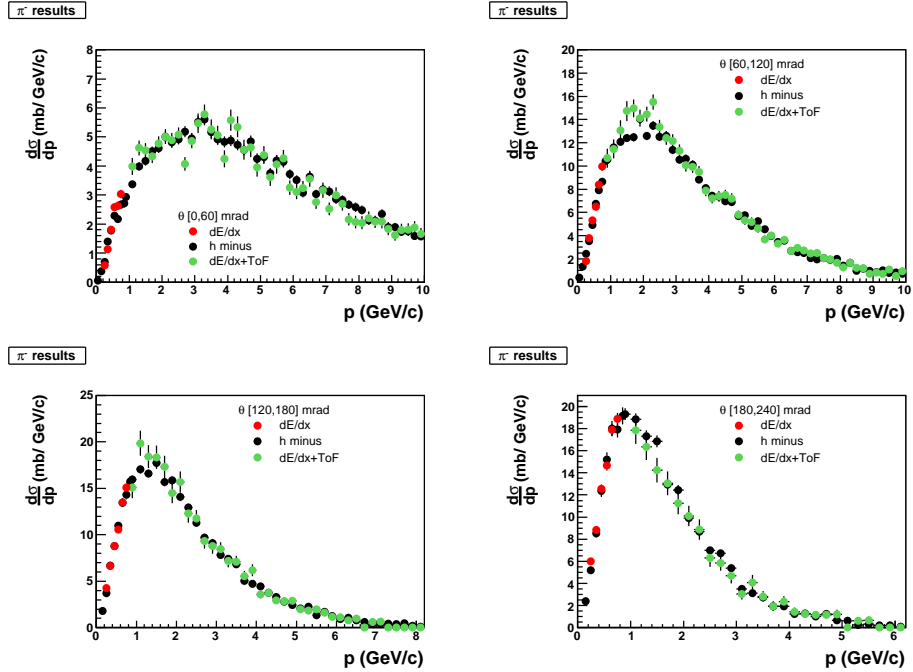, width=0.8\textwidth}
\end{center}
\caption{Differential cross-section for negative pions in four different angular
  bins (mentioned on plots). Markers correspond to different analysis procedures.}
\label{fig:xsec-pim}
\end{figure}

\begin{figure}[!h]
\begin{center}
\epsfig{figure=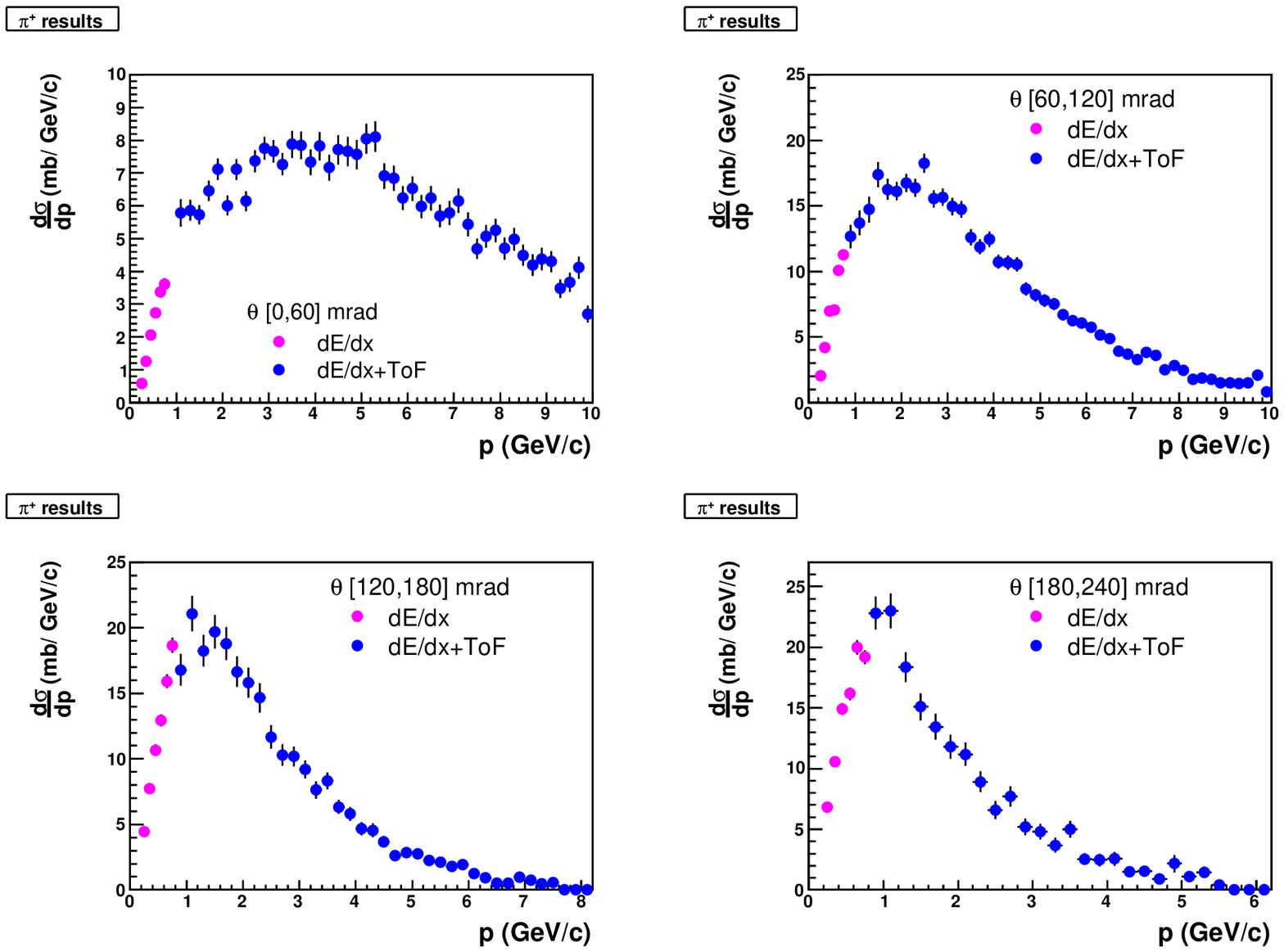, width=0.8\textwidth}
\end{center}
\caption{Differential cross-section for positive pions in four different angular
  bins (mentioned on plots). Markers correspond to different analysis procedures.}
\label{fig:xsec-pip}
\end{figure}

The proton on Carbon data at 31 GeV/c from the 2007 pilot run have
been used to produce preliminary spectra of both negative (up to 15
GeV/c momentum) and positive (up to 10 GeV/c momentum) pions in
angular bins of 60 mrad. Differential cross-sections for different
angular bins are shown in Fig.~\ref{fig:xsec-pim} and
Fig.~\ref{fig:xsec-pip} respectively. Only statistical errors are
shown, while results are still quoted with 20\% systematic errors
coming from the current level of disagreement obtained when comparing results from
different analysis procedures in some bins.\\
Three procedures have been developed for the analysis: the negative
hadron analysis, the $dE/dx$ analysis below 1 GeV/c momentum and the
combined $ToF$-$dE/dx$ analysis starting from 0.8 GeV/c, which is
necessary for the $\pi^+$ spectra. The three procedures give
consistent results within the quoted systematic errors for the
negative pion analysis (see Fig.~\ref{fig:xsec-pim}) and continuity is
observed between the
procedures used for the positive pion analysis (see Fig.~\ref{fig:xsec-pip}).\\
The thin Carbon target results also include the determination of the
absolute inelastic cross-section (used for normalization) of proton on
Carbon at 31 GeV/c, and preliminary comparisons with different models
such as GiBUU~\cite{GiBUU}, Geant4~\cite{GEANT4} and
FLUKA-standalone. Work is currently performed to lower the quoted
systematic errors.

\section{NA61 data for the T2K neutrino flux predictions}
\begin{figure}[!h]
\begin{center}
\epsfig{figure=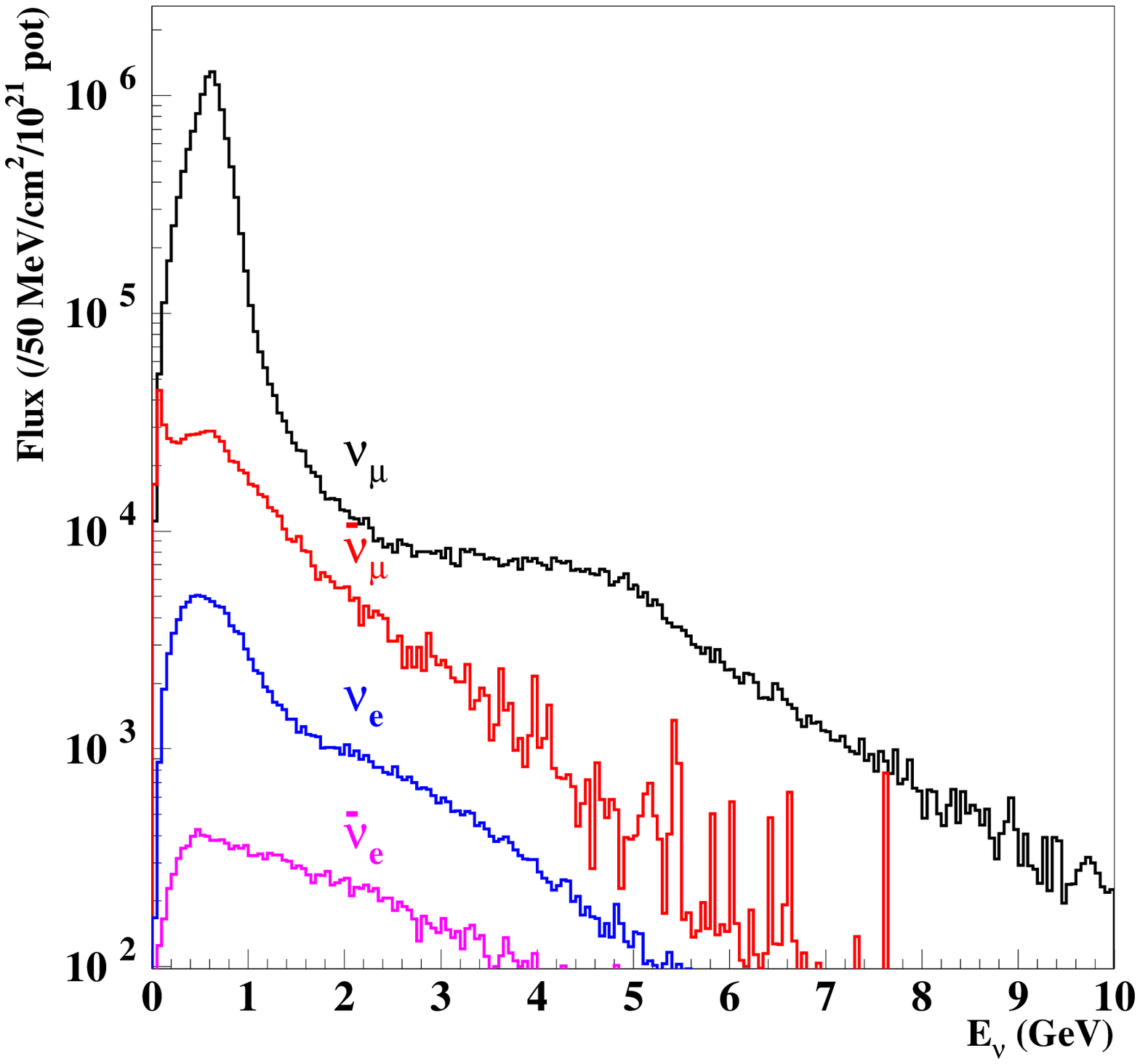, width=0.32\textwidth}
\epsfig{figure=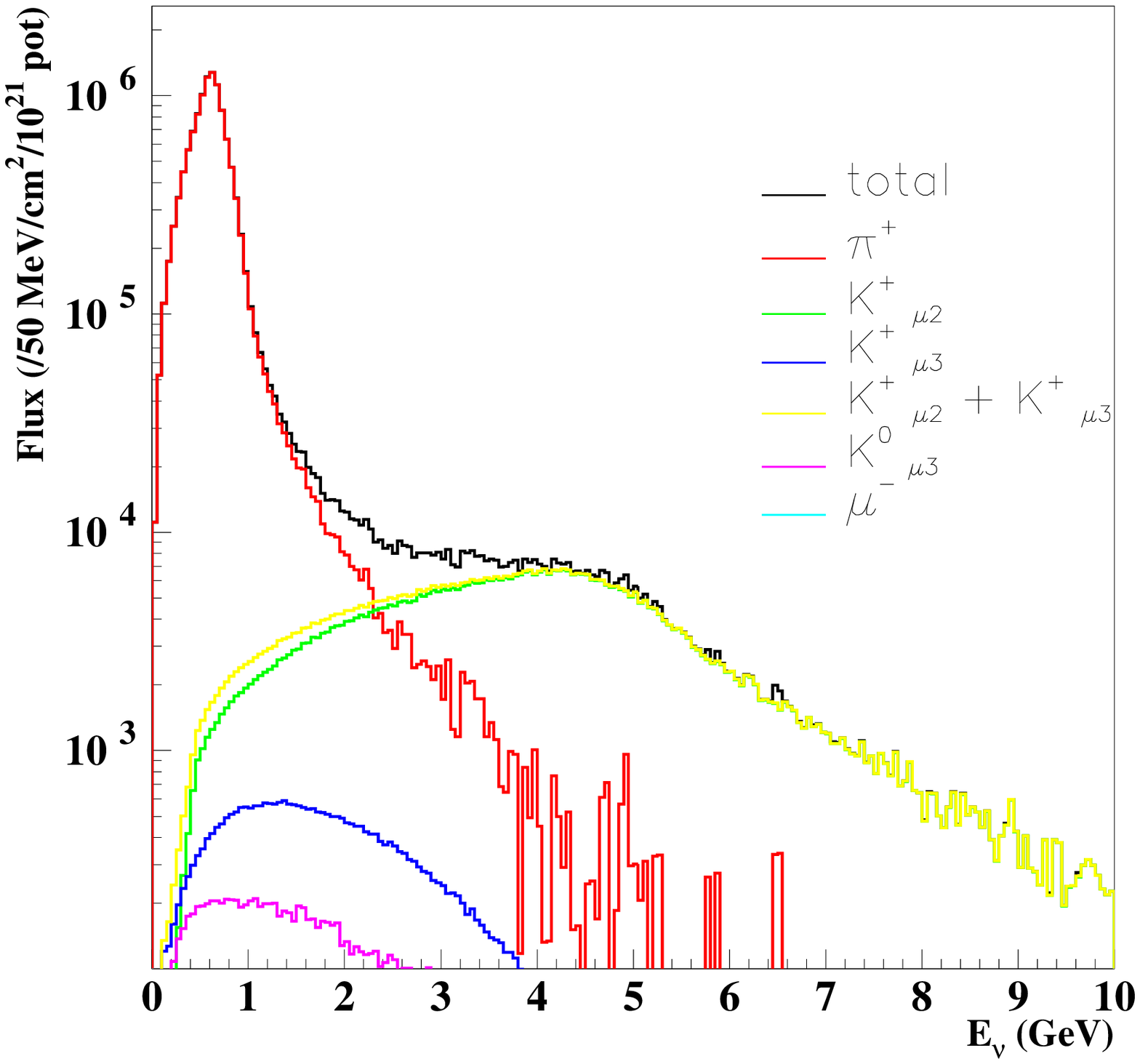, width=0.32\textwidth}
\epsfig{figure=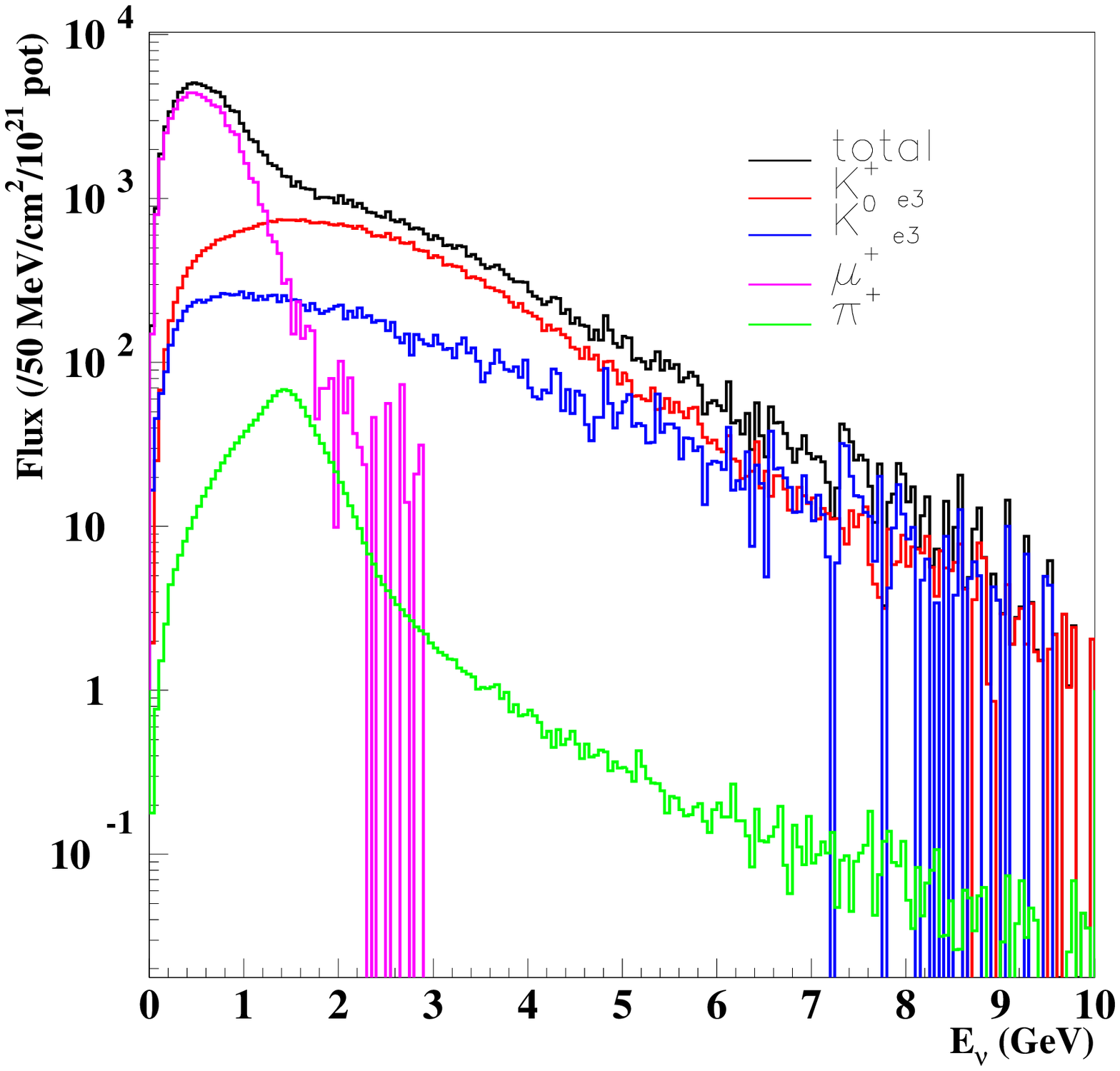, width=0.32\textwidth}
\end{center}
\caption{Neutrino fluxes for all species ($\nu_{\mu}$,
  $\bar{\nu}_{\mu}$, $\nu_e$ and $\bar{\nu}_e$) at the far detector
  (left). Parent contributions to $\nu_{\mu}$ (middle) and $\nu_e$
  (right) fluxes at the far detector.}
\label{fig:fluxes-sk}
\end{figure}

JNUBEAM (release 10a) is the T2K beam simulation~\cite{T2K} program. It
has been used to predict fluxes for the four different neutrino
species ($\nu_{\mu}$, $\bar{\nu}_{\mu}$, $\nu_e$ and $\bar{\nu}_e$) at
both T2K near and far detectors. Fig.~\ref{fig:fluxes-sk} shows total
fluxes for all species at the far detector, as well as contributions
from different parent particles for $\nu_{\mu}$ and $\nu_e$ species.

\begin{figure}[!h]
\begin{center}
\epsfig{figure=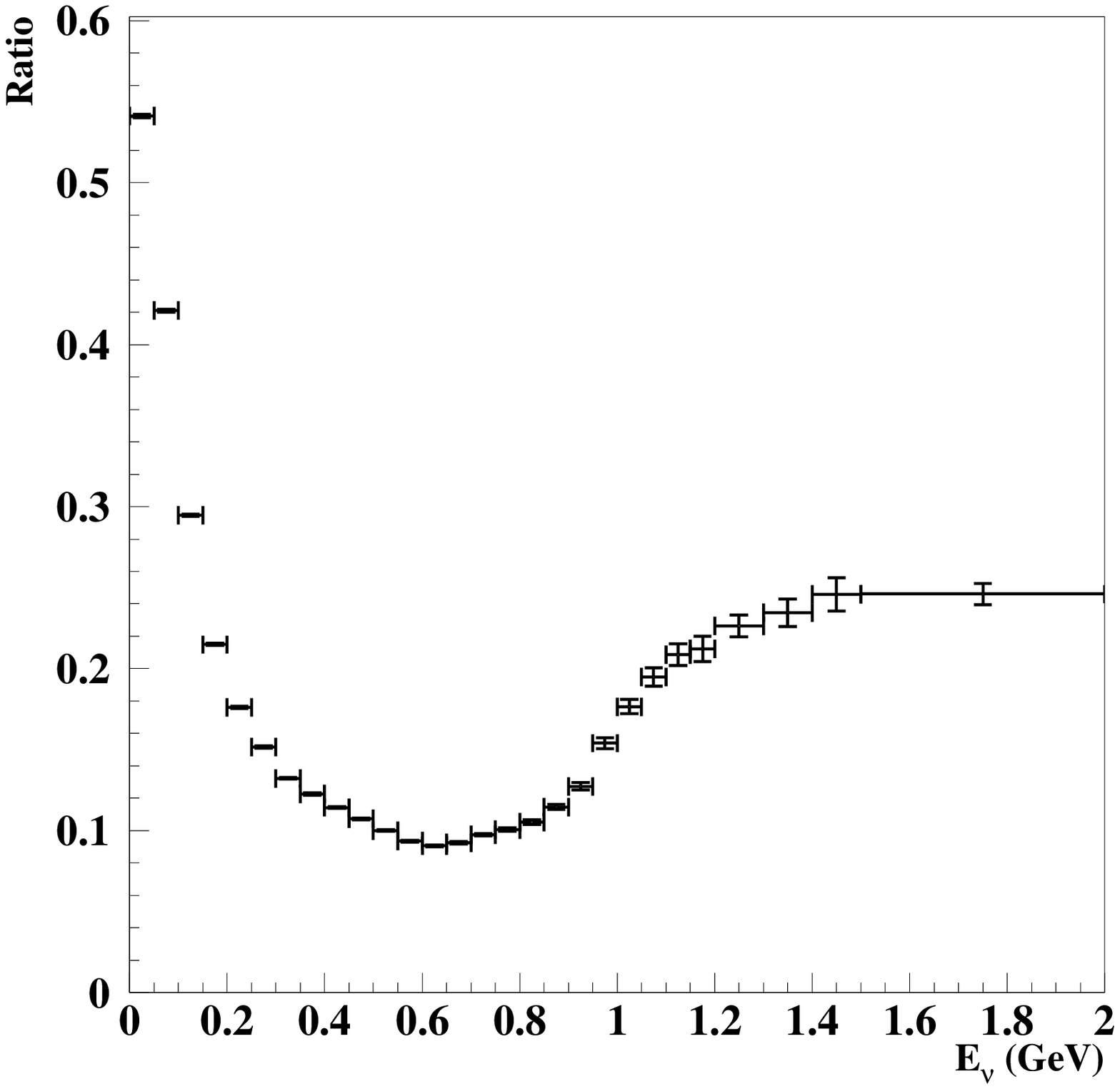, width=0.45\textwidth}
\epsfig{figure=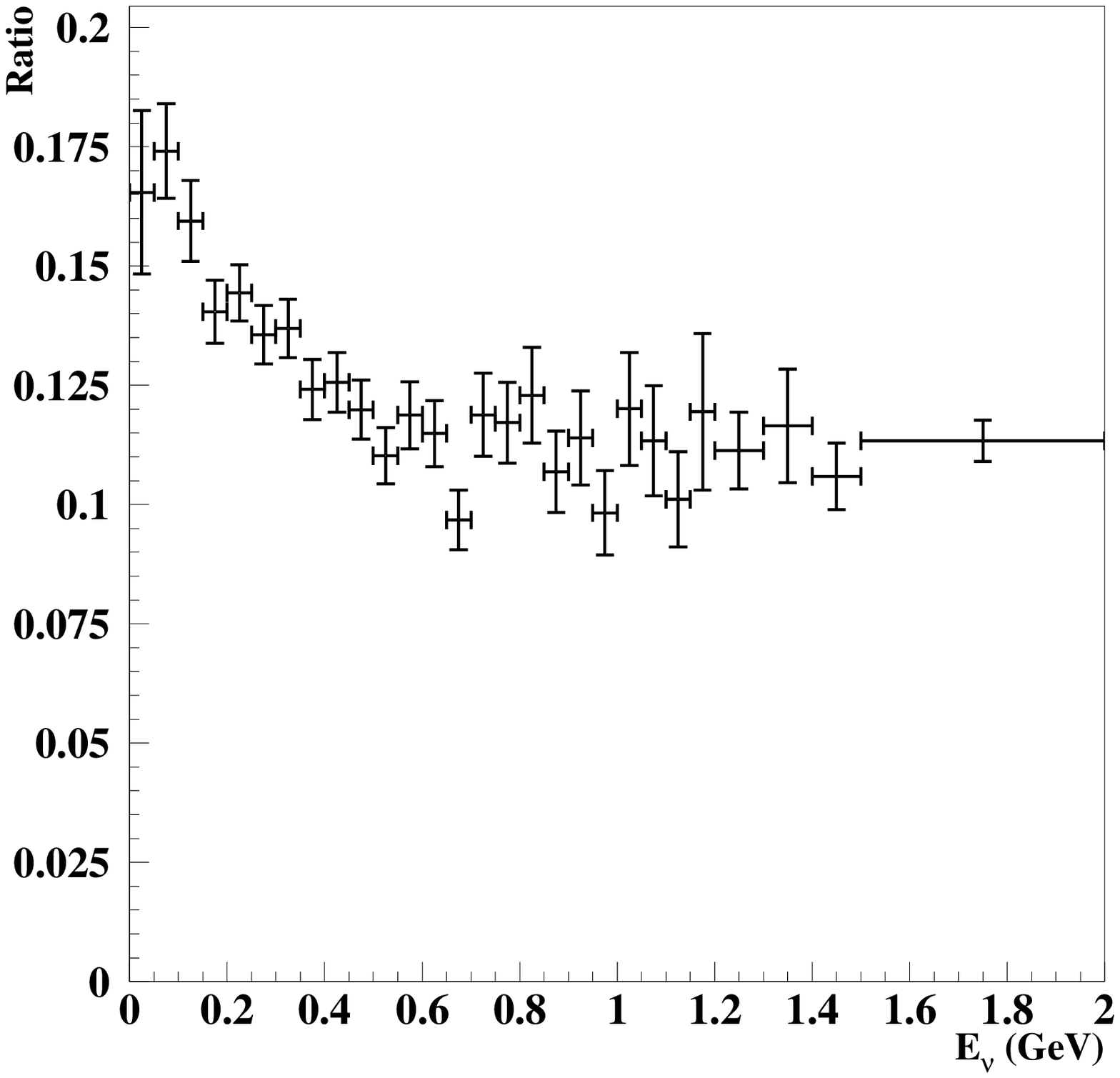, width=0.45\textwidth}
\end{center}
\caption{Ratio of the out-of-target contribution to the total
  contribution for $\nu_{\mu}$ (left) and $\nu_e$ (right) components at the far
detector.}
\label{fig:ratios}
\end{figure}

Contributions to the neutrino fluxes have been defined according to
the NA61 measurements.
\textit{In-target} and \textit{out-of-target} contributions refer to
the measurements with the full size T2K replica target.
\textit{Indirect} and \textit{direct} contributions refer to the
measurements with the thin target in which only primary interactions
are measured.\\
The \textit{in-target} contribution comes from neutrino
parents produced inside the target (from primary or secondary
interactions), as well as from neutrino parents (such as muons)
produced in the decay of particles originating from the target
($\sim$5-10 \% of the contribution). Apart from decays out of the
target, this contribution corresponds to what is measured with the
long target.
The \textit{out-of-target} contribution
accounts for neutrino parents produced in re-interactions in the other
elements of the beam line (magnetic horns in particular).\\
The \textit{direct} contribution comes from neutrino parents produced
in the primary interaction, as well as from parents produced in the
decay of those secondary particles. Apart from decays, this
contributions corresponds to what is measured with the thin target.
The \textit{indirect} contribution accounts for parents produced in
secondary interactions in and out of the target.\\
As shown in Fig.~\ref{fig:ratios}, the ratio of the out-of-target
contribution to the total contribution is $\sim$10 \% at peak energy
for both $\nu_{\mu}$ and $\nu_e$ species. The equivalent ratio between 
indirect and total contributions is $\sim$40 \% at peak energy for both
$\nu_{\mu}$ and $\nu_e$.  This conclusion stresses the importance of
the replica target measurements: providing tracks with momentum and
angle (with respect to the beam direction) at exit point on the target
surface will allow to predict directly a fraction of the neutrino flux at both
near and far detectors as high as $\sim$90 \% for both $\nu_{\mu}$ and
$\nu_e$ components.

Studies showed variations (both in shape and normalization) of the
absolute neutrino fluxes as a function of the neutrino parent exiting
point position on the target surface, as well as of the beam profile
used in the simulation. Those considerations lead to a binning of the
replica target data consisting in six equidistant longitudinal bins
and three to four radial bins.\\ 
The NA61-SHINE thin target measurements can provide pion and kaon
production cross-sections as direct input to the T2K beam
simulation. In this case, still $\sim$40 \% of the neutrino fluxes would
require using models for secondary interactions.
Due to the limited azimuthal acceptance of NA61-SHINE, the replica target data cannot
be used as a direct input to the simulation on an event-by-event
basis. However, they can be used to re-weight the beam Monte-Carlo
using the event generators of the T2K beam simulation within the
NA61-SHINE simulation chain. In this case, ~10 \% of the neutrino
fluxes would still require using models to predict secondary
interactions outside the target. A method has been developed to propagate
those re-weighting factors (and associated statistical and systematic
errors) from the \{$p,\theta$\} phase space (in longitudinal and
radial bins) of relevance in T2K, to the neutrino flux predictions. 


\end{document}